\documentstyle[11pt,aaspp4,epsf]{article}

\def\nh{n_{H,22}}
\def\einst{{\sl Einstein}}
\def\ros{{\sl ROSAT}}
\def\chan{{\sl Chandra}}
\begin{document}
\lefthead{Pavlov, Zavlin, Aschenbach, et al.}
\righthead{Compact Central Object in Cas A}
\vspace*{-25mm}
\hfill Submitted to {\it The Astrohysical Journal}  
\vspace*{15mm}
\title{The Compact Central Object in Cas A: A Neutron Star
with Hot Polar Caps or a Black Hole?}
\author{G.~G.~Pavlov\footnote{
The Pennsylvania State University, 525 Davey Lab,
University Park, PA 16802, USA; pavlov@astro.psu.edu}, 
V.~E.~Zavlin\footnote{
Max-Planck-Institut f\"ur Extraterrestrische Physik, D-85740
Garching, Germany; zavlin@xray.mpe.mpg.de},
B.~Aschenbach$^2$, J.~Tr\"umper$^2$, and D.~Sanwal$^1$}
\begin{abstract}
The central pointlike X-ray source of the Cas A 
supernova remnant was discovered in the \chan\ First Light Observation
and found later in the archival \ros\ and \einst\ images.
The analysis of 
these data
does not show statistically significant variability of the source.
Because of the small number of photons detected, 
different spectral models can fit the observed spectrum.
The power-law fit yields 
the photon index $\gamma=2.6$--4.1,
and luminosity $L(0.1$--$5.0~{\rm keV})=
(2$--$60)\times 10^{34}$~erg~s$^{-1}$, for $d=3.4$~kpc.
The power-law 
index is 
higher, 
and the 
luminosity lower,
than those observed from very young pulsars.
One can fit the spectrum equally well 
with a blackbody model
with 
$T=6$--8~MK, $R=0.2$--$0.5$~km, 
$L_{\rm bol}=(1.4$--$1.9)\times 10^{33}$~erg~s$^{-1}$.
The inferred radii are too small, and the temperatures
too high, for the radiation could be interpreted as emitted from
the whole surface of a uniformly heated neutron star.
Fits with the neutron star atmosphere models increase the 
radius and reduce the temperature, but these parameters are
still substantially different from those expected for a young neutron
star. One cannot exclude, however, that the
observed emission originates from hot spots on a cooler
neutron star surface. Because of strong interstellar absorption, 
the possible low-temperature component
gives a small contribution to the observed spectrum; an upper limit
on the (gravitationally redshifted) surface 
temperature is $T_s^\infty < 1.9$--2.3~MK,
depending on chemical composition of the surface and star's radius.
Amongst several possible interpretations, 
we favor a model of a strongly magnetized neutron star
with magnetically confined hydrogen or helium polar caps
($T_{\rm pc}^\infty \approx 2.8$~MK, $R_{\rm pc}\approx
1$~km) on a cooler iron surface ($T_{\rm s}^\infty \approx 1.7$ MK).
Such temperatures are consistent with the 
standard models of neutron star cooling.
Alternatively, the observed radiation may be interpreted as emitted
by a compact object (more likely, a black hole) accreting from
a fossil disk or from a late-type dwarf in a close binary.
\end{abstract}
\keywords{stars: neutron --- supernovae: individual (Cas A) --- X-rays: stars}
\section{Introduction}
Cassiopeia A is the brightest shell-type galactic supernova remnant (SNR)
in X-rays and radio, and the youngest SNR observed
in our Galaxy. The radius of the approximately spherical shell
is $\approx 2'$, which corresponds to $\approx 2$ pc for
the distance $d=3.4^{+0.3}_{-0.1}$~kpc (Reed et al.~1995).
The supernova which gave rise to Cas A was probably first observed
in 1680 (Ashworth 1980).
It is thought to be a Type II supernova
caused by explosion of a very massive Wolf-Rayet
star (Fesen, Becker \& Blair 1987). 
Optical observations of Cas A show
numerous oxygen-rich fast-moving knots (FMK),
with  velocities of about 5000~km~s$^{-1}$, and
slow-moving quasi-stationary flocculi,
with typical velocities of about 200~km~s$^{-1}$, which emit
H$\alpha$ and strong lines of nitrogen. X-ray observations of Cas A
show numerous clumps of hot matter emitting strong
Si, S, Fe, Ar, Ne, Mg and Ca lines 
(Holt et al.~1994, and references therein). 
Because this SNR lies at
the far side of the Perseus arm, with its patchy distribution
of the interstellar gas, the interstellar absorption
varies considerably across the Cas A image
(e.g., Keohane, Rudnick \& Anderson 1996). Numerous radio, optical
and X-ray measurements of the hydrogen column density
(e.g., Schwarz, Goss \& Kalberla 1997; Hufford \& Fesen 1996;
Jansen et al.~1988; Favata et al.~1997)
show a strong scatter within a range $0.5\la \nh\la 1.6$, where 
$\nh\equiv n_H/(10^{22}~{\rm cm}^{-2})$. Based on recent results,
we consider $0.8\la\nh\la 1.4$ as plausible values
for the central region of the Cas A image.

In spite of considerable efforts to detect 
a compact remnant of the supernova explosion 
only upper limits on its flux had been established at different
wavelengths until 
a pointlike X-ray source
was discovered 
close to the Cas A center (Tananbaum et al.~1999) in the First Light
Observation with the \chan\ X-ray Observatory (see Weisskopf et al.~1996
for a description). After this discovery,
the same source was found in the \ros\ HRI image of 1995--96
(Aschenbach 1999) and \einst\ HRI images of 1979 and 1981 (Pavlov \&
Zavlin 1999). 

In this Letter we present the first analysis on the central source
spectrum observed with \chan\ (\S 2), together with the analysis
of the \ros, \einst, and {\sl ASCA} observations (\S 3). 
Various interpretations
of these observations are discussed in \S 4.
\section{\chan\ ACIS observation and the point source spectrum}

The SNR Cas A was observed several times 
during the \chan\ Orbital Activation and Calibration Phase. 
For our analysis, we chose four observations of 1999 August 20--23
with the S array of the
Advanced CCD Imaging Spectrometer (ACIS; Garmire 1997). 
In these observations Cas A was imaged on the backside-illuminated chip S3.
The spectral response of this chip is presently known better that those of the 
frontside-illuminated chips used in
a few other ACIS observations of Cas A. 
We used the processed 
data products available from the public \chan\ Data Archive.
The observations were performed in the Timed Exposure mode, with 
a frame integration time of 3.24~s. The durations of the observations were
5.03, 2.04, 1.76, and 1.77~ks.
Because of telemetry saturation, 
the effective exposures were 2.81, 1.22, 1.06, and 1.05 ks,
respectively. Since the available ACIS response matrices
were generated for the set of grades G02346,
we selected events with these grades. Events with pulse height amplitudes
exceeding 4095 ADU ($\approx 0.7\%$ of the total number) were discarded as
generated by cosmic rays.
The images of the pointlike source look slightly elongated,
but this elongation is likely caused by
errors in the aspect solution, and the overall shapes
of the images is consistent with the assumption that this is
a point source. Its positions in the four observations are consistent
with that reported by Tananbaum et al.~(1999):
$\alpha_{2000}=23^{\rm h}23^{\rm m}27\fs 94$,
$\delta_{2000}=+58^\circ 48' 42\farcs 4$.
 For each of the images, we extracted 
the source+background counts from 
a $3''$ radius circle around the point source center,
and the background from an elliptical region around the circle, with an 
area of about 10 times that of the circle. 
After subtracting the background,
we obtained the source countrates
$112\pm 8$, $121\pm 13$, $112\pm 14$, and $127\pm 15$ ks$^{-1}$
(counts per kilosecond). The countrate values and the light curves
are consistent with the assumption 
that the source flux remained 
constant during the 4 days, with the countrate of 
$116\pm 6$ ks$^{-1}$.

 For the analysis of the point source spectrum, we chose the longest
of the ACIS-S3 observations.
We grouped the pulse-height spectrum for 306 source counts
into 14 bins in the 0.8--5.0~keV range
(Fig.~1). Each bin has more than 20 counts (except for the highest-energy bin 
with 8 counts).
The spectral fits were performed with the XSPEC package.

If the source is an active
pulsar, we can expect that its X-ray radiation is emitted by relativistic
particles and has a power-law spectrum.
The power-law fit (upper panel of Fig.~2)
yields a photon index $\gamma=3.2^{+0.9}_{-0.6}$ 
(all uncertainties are given at a $1$-$\sigma$ confidence level) 
that is considerably larger than 
$\gamma=1.4$--2.1
observed for X-ray radiation from youngest pulsars (Becker \& Tr\"umper 1997).
The hydrogen column density, 
$\nh=1.7^{+0.7}_{-0.5}$,
inferred from the power-law fit somewhat exceeds estimates
obtained from independent measurements (see \S 1).
The (unabsorbed) X-ray luminosity in the 0.1--5.0~keV range,
$L_X=5.8^{+58.3}_{-4.3}\times 10^{34}~d_{3.4}^2$~erg~s$^{-1}$,
where $d_{3.4}=d/(3.4~{\rm kpc})$, is lower than those
observed from very young pulsars (e.g., 
$1.5\times 10^{36}$ and $2.3\times 10^{36}$~erg~s$^{-1}$ 
for the Crab pulsar and PSR B0540--69, in the same energy range).

If the source is a neutron star (NS), but not an active pulsar, thermal
radiation from the NS surface can be observed.
The blackbody fit (middle panel of Fig.~2)
yields a temperature $T_{\rm bb}^\infty=7.1^{+1.1}_{-1.0}$~MK
and a sphere radius 
$R_{\rm bb}^\infty =0.29^{+0.16}_{-0.09}~d_{3.4}$~km, 
which correspond to a bolometric luminosity 
$L_{\rm bb,bol}^\infty =
1.6^{+0.3}_{-0.2}\times 10^{33}~d_{3.4}^2$~erg~s$^{-1}$.
(We use the superscript $^\infty$ to denote the observed
quantities, distinguishing them
from those at the NS surface: $T^\infty=g_rT$, $L^\infty=
g_r^2L$, $R^\infty =g_r^{-1}R$, where $g_r=[1-2GM/Rc^2]^{1/2}=
[1-0.41 M_{1.4}R_6^{-1}]^{1/2}$
is the gravitational redshift factor, $M= 1.4 M_{1.4} M_\odot$ and
$R=10^6 R_6$~cm are the NS mass and radius).
The temperature is too high,
and the radius is too small, to interpret the detected X-rays
as emitted from the whole surface of a cooling NS
with a uniform temperature distribution.
The inferred hydrogen column density, $\nh=0.6^{+0.5}_{-0.3}$,
is on a lower side of the plausible $n_H$ range.

Since fitting observed X-ray spectra with light-element
NS atmosphere models yields lower effective temperatures and
larger emitting areas (e.~g., Zavlin, Pavlov \& Tr\"umper 1998),
we fit the spectrum with a number of
hydrogen and helium NS atmosphere models
(Pavlov et al.~1995; Zavlin, Pavlov \& Shibanov 1996), for several values
of NS magnetic field. These fits show that
the assumption that the observed radiation is emitted from the
whole surface of a 10-km radius NS with a uniform temperature
still leads to unrealistically large distances, 
$\sim 20$--50 kpc.
Thus, both the blackbody fit and H/He atmosphere fits 
hint that, if the object is a NS, 
the observed radiation emerges from hot spots on its surface
(see \S 4).
An example of such a fit, for polar caps covered with a hydrogen atmosphere
with $B=5\times 10^{12}$~G, 
is shown in the bottom panel of Figure~2. The model spectra used in this
fit were obtained 
assuming
the NS 
to be an orthogonal rotator (the angles
$\alpha$, between the magnetic and rotation axes, and $\zeta$, between
the rotation axis and line of sight, equal $90^\circ$).
The inferred effective 
temperature of the caps is
$T_{\rm pc}=5.9^{+1.4}_{-1.6}$~MK
(which corresponds to 
$T_{\rm pc}^\infty=4.5^{+1.1}_{-1.2}$~MK),
the polar cap radius 
$R_{\rm pc}=0.8^{+1.1}_{-0.3}~d_{3.4}$~km,
and $\nh=0.7^{+0.4}_{-0.3}$.
The bolometric luminosity of two polar caps is
$L_{\rm pc,bol}=2.6^{+1.7}_{-0.5}\times 10^{33}~d_{3.4}^2$~erg~s$^{-1}$.
The temperature $T_{\rm pc}$
can be 
lowered, and the polar cap radius increased,
if we see the spot face-on during the
most part of the period --- extreme values, $T_{\rm pc}=3.8^{+1.1}_{-0.8}$~MK
($T_{\rm pc}^\infty =2.9^{+0.8}_{-0.5}$~MK) and 
$R_{\rm pc}=0.9^{+1.2}_{-0.3}~d_{3.4}$~km, 
at $\nh =0.8^{+0.5}_{-0.4}$, correspond to $\alpha=\zeta=0$.

The fits with the one-component thermal models
implicitly assume that the temperature of the rest 
of the NS surface 
is so low that its radiation is not seen by ACIS. On the other hand,
according to the NS cooling models (e.g., Tsuruta 1998),
one should expect that, at the age of 320 yr,
the (redshifted) surface temperature 
can be as high as 2~MK for the so-called standard cooling
(and much lower, down to 0.3 MK, for accelerated cooling). 
To constrain the temperature
outside the polar caps, 
we repeated the polar
cap fits with the second thermal component added, at a fixed NS radius
and different (fixed) values of surface temperature $T_{\rm s}$.
With this approach we estimated upper limits on the lower 
temperature, $T_{\rm s}^\infty < 1.9$--2.3~MK (at a 99\%
confidence level), depending on
the low-temperature model chosen. These fits show that the model
parameters are strongly correlated --- 
the increase of $T_{\rm s}$ shifts the best-fit $T_{\rm pc}$ downward,
and $n_{H}$ upward. 
For example, using an iron atmosphere model for the low-temperature
component and assuming a hydrogen polar cap, we obtain an
acceptable fit (see Fig.~1)
for $T_{\rm s}^\infty = 1.7$~MK, $R=10$~km, 
$T_{\rm pc}^\infty =2.8$~MK,
$R_{\rm pc}=1.0~d_{3.4}$~km, 
$\nh = 1.1$. 
Note that this $T_{\rm s}$ is consistent with the predictions
of the standard cooling models,
and $n_H$ is close to most plausible values
adopted for the central region of the SNR.
\section{Analysis of the \ros, \einst\ and {\sl ASCA} images}
We reanalyzed the archival data on Cas A
obtained during a long \ros\ HRI observation, 
between 1995 December~23 and 1996 February~1
(dead-time corrected exposure 175.6~ks). 
The image shows a pointlike central source at the position
$\alpha_{2000}=23^{\rm h}23^{\rm m}27\fs 57$,
$\delta_{2000}=58^\circ 48\arcmin 44\farcs 0$
(coordinates of the center of the brightest $0\farcs 5\times 0\farcs 5$ pixel),
consistent with that reported by Aschenbach (1999).
Its separation from the \chan\ point source
position, $3\farcs 3$, is smaller
than the \ros\ absolute pointing uncertainty (about $6''$; Briel et al.~1997).
Measuring the source countrate is complicated by the spatially
nonuniform background.
Another complication is that the 40-day-long exposure actually
consists of many single exposures of very different durations.
Because of the absolute pointing errors,
combining many single images in one leads to additional 
broadening of the point source function (PSF). 
To account for these complications, we used several apertures (with
radii from $3''$ to $7''$) for source+background extraction,
measured background in several regions with visually the same
intensity as around the source, discarded short single exposures,
and used various combinations of long single exposures for countrate
calculations.
This analysis yields a source countrate
of $4.6\pm 0.8$ ks$^{-1}$ 
(corrected for the finite apertures).

We also re-investigated the archival data on Cas A 
obtained with the \einst\ HRI in observations
of 1979 February 8 
(42.5~ks exposure) 
and 1981 January 22--23 
(25.6~ks exposure). In each of the data sets 
there is a pointlike source
at the positions 
$\alpha_{2000}=23^{\rm h}23^{\rm m}27\fs 83$,
$\delta_{2000}=58^\circ 48\arcmin 43\farcs 9$,
and $\alpha_{2000}=23^{\rm h}23^{\rm m}27\fs 89$,
$\delta_{2000}=58^\circ 48\arcmin 43\farcs 7$, respectively,
consistent with those reported by Pavlov \& Zavlin (1999). 
The separations from the \chan\ position, 
$1\farcs 7$ and $1\farcs 4$, and from the \ros\ position, 
$2\farcs 0$ and $2\farcs 5$, 
are smaller than the nominal absolute position uncertainty
($\sim 4''$ for \einst). 
Since the observations were short, estimating the source countrates 
is less complicated than for the \ros\ HRI observation.
The source+background counts were selected from $5''$-radius
circles, and the background was measured from annuli of $10''$ outer radii
surrounding the circles. 
The source countrates, calculated with account for the HRI PSF
(Harris et al.~1984), are
$2.2\pm 0.5$ (Feb 1979), $2.7\pm 0.8$~ks$^{-1}$ (Jan 1981),
and $2.4\pm 0.6$~ks$^{-1}$ (for the combined data).
The countrate is consistent with the 
upper limit of $7.5$ ks$^{-1}$,
derived by Murray et al.~(1979) from the longer of the two 
observations. 

To check whether the source radiation varied during the two decades,
we plotted the lines of constant
\ros\ and \einst\ HRI
countrates in Figure 2. 
 For all the three one-component models, 
the domains of model parameters corresponding to
the \einst\ HRI countrates within a $\pm 1\sigma$ range
are broader than the 99\%
confidence domains obtained from the \chan\ spectra.
The 1-$\sigma$ domains corresponding to the \ros\ HRI countrate 
overlap with the 1-$\sigma$ confidence regions
obtained from the spectral data.
Thus, the source countrates detected with
the three instruments do not show
statistically significant variability of the source.

We also examined numerous archival {\sl ASCA} observations of Cas A
(1993--1999)
and failed to detect the central point source
on the high background produced by bright SNR structures smeared by
poor angular resolution of the {\sl ASCA} telescopes.
In the longest of the {\sl ASCA} SIS observations
(1994 July 29; 15.1 ks exposure) the point source
would be detected at a 3-$\sigma$ level if its flux were
a factor of 8 higher 
than that observed with \chan, \ros, and \einst.
The {\sl ASCA} observations show that there were no strong outbursts of
the central source.
\section{Discussion}
The observed 
X-ray energy
flux, $F_X$, of the compact central object 
(CCO\footnote{According to the convention recommended by the
{\sl Chandra} Science Center, this source should be named   
CXO~J232327.9+584842. We use the abbreviation CCO for brevity.}) 
is 3.6, 6.5, and $8.2\times
10^{-13}$~erg~cm$^{-2}$~s$^{-1}$ in 0.3--2.4, 0.3--4.0 and 0.3--6.0~keV
ranges, respectively.
Upper limits on its optical-IR fluxes,
$F_R\la 3\times 10^{-15}$~erg~cm$^{-2}$~s$^{-1}$
and $F_I\la 1\times 10^{-14}$~erg~cm$^{-2}$~s$^{-1}$,
can be estimated from the magnitude limits, $R\ga 24.8$ and
$I\ga 23.5$, found by van den Bergh \& Pritchet (1986).
This gives, e.g., $F_X/F_R \ga 100$ for the \ros\ energy range,
and $F_X/F_R \ga 200$ for the \chan\ and \einst\ ranges.
The flux ratios are high enough to exclude coronal emission from
a noncompact star as the source of the observed X-ray radiation. 
A hypothesis that CCO is a background AGN or a cataclismic
variable cannot be completely rejected, but its probability looks
extremely low, given the high X-ray-to-optical flux ratio,
the softness of the spectrum, and the lack of indications on variability. 

The strong argument for CCO to be a compact remnant
of the Cas A explosion is its proximity to the Cas A center.
In particular, this source lies $7''$--$11''$ south of the
SNR geometrical center determined from the radio image
of Cas A (see Reed et al.~1995, and references therein).
The source separation, $1''$--$5''$, from the SNR expansion center, found by
van den Bergh \& Kamper (1983) from the analysis of proper
motions of FMKs, corresponds to 
a transverse velocity of
50--250 km~s$^{-1}$ (for $d=3.4$ kpc, $\tau=320$ yr).
Much higher transverse velocities, 800--1000 km~s$^{-1}$,
correspond to the separation, $16''$--$20''$, from the position
of the apparent center of expanding SNR shell
derived by Reed et al.~(1995) from the radial velocities of FMKs. 
Thus, if CCO is the compact remnant of the SN explosion, it is
moving south (or SSE) from the Cas A center with a
transverse velocity of a few hundred km~s$^{-1}$, 
common for radio pulsars.

If CCO is an isolated (nonaccreting) object, it 
might be an active pulsar with an unfavorable orientation
of the radio beam (a limit on 
the pulsed flux of 80 mJy at 408 MHz
was reported by Woan \& Duffett-Smith 1993).  
However, a lack of a plerion or a resolved synchrotron
nebula, together with the steep X-ray spectrum and 
low luminosity 
(see \S 2) do not support this hypothesis. 
The lack of the pulsar activity has been found in several X-ray 
sources associated with young
compact remnants of SN explosions (e.g., Gotthelf, Vasisht \& Dotani 1999);
it may be tentatively explained 
by superstrong ($\ga 10^{14}$~G) magnetic fields which may suppress
the one-photon pair creation in the pulsar's 
acceleration gaps (Baring \& Harding 1998).

If CCO is an isolated NS without pulsar activity,
one may assume that
the observed X-rays are emitted from the NS surface.
In this case, we also have to assume an intrinsically
nonuniform surface temperature distribution
to explain the small size and high temperature
of the emission region.
Slight nonuniformity of the surface temperature
can be caused by anisotropy of heat conduction
in the strongly magnetized NS crust (Greenstein \& Hartke 1983).
However, this nonuniformity is not strong
enough to explain the small apparent areas of the emitting regions.
Some nonuniformity might be expected in magnetars, 
if they are indeed powered by decay of their superstrong
magnetic fields (Thompson \& Duncan 1996; Heyl \& Kulkarni 1998) and
a substantial fraction of the thermal energy is produced in the outer
NS crust. In this case, 
the hotter regions of the NS surface would be those
with stronger magnetic fields. 
If additional investigations will demonstrate 
quantitatively that the observed 
luminosity of $\sim 10^{33}$ erg~s$^{-1}$ can be emitted from a small
fraction, $\sim 10^{-2}$, of the magnetar's surface, we should
expect that the radiation is pulsed, with a probable period of a few seconds
typical for magnetars.  We can also speculate that 
CCO is a predecessor of a soft gamma-repeator
(such a hypothesis has been proposed by Gotthelf et al.~1999 for the
central source of the Kes 73, which shows a spectrum similar to CCO, 
albeit emitted from a larger area).

Higher temperatures of polar caps can be explained by
different chemical compositions
of the caps and the rest of the NS surface.
Light-element polar caps could form just after the SN
explosion via fallback of a fraction of the ejected matter onto
the magnetic poles.
Due to fast stratification in the strong gravitational field,
the upper layers of the polar caps will be comprised of the lightest
element present.
The thermal conductivity in the liquid portion of thin
degenerate NS envelopes, which is responsible for
the temperature drop from the nearly isothermal interior to the surface,
is proportional to $Z^{-1}$, where $Z$ is the ion charge
(Yakovlev \& Urpin 1980). This means that low-$Z$ envelopes are 
more efficient heat conductors than
high-$Z$ ones, so that a light-element (H, He) surface has a higher
effective temperature for a given temperature $T_b$ at the outer
boundary of the internal isothermal region. Approximately, the
effective surface temperature is proportional to $Z^{-1/4}$ if
the chemical composition of the envelope does not vary with depth,
so that the surface of a hydrogen envelope can be $2.2$ times
hotter than that of an iron envelope.
Numerical calculations of Chabrier, Potekhin \& Yakovlev (1997)
give a smaller factor (1.6--1.7 for temperatures of interest), 
with account for burning of light elements into heavier ones in the 
hot bottom layers of the envelope, 
but neglecting the effects of 
strong magnetic fields which can somewhat increase this factor (Heyl \&
Hernquist 1997). 
Hence, the light-element cap should be hotter than the rest of the NS surface.
For instance, for $T_b=400$~MK, the effective
temperatures of the H cap and Fe surface are $T_{\rm pc}^\infty = 2.8$ MK
and $T_{\rm s}^\infty=1.7$ MK, for $M=1.4 M_\odot$, $R=10$ km.
As we have shown in \S 2, a two-component model spectrum with such
temperatures is consistent with the observed CCO spectrum, for
$R_{\rm pc}\simeq 1$~km. The thickness of the hydrogen cap,
$\ga 2\times 10^{13}$ g~cm$^{-2}$, needed to provide such
a temperature difference, 
corresponds to the total cap mass 
$M_{\rm pc}\sim 6\times 10^{23} (R_{\rm pc}/1~{\rm km})^2~{\rm g}$.
 For lower $M_{\rm pc}$, the temperature difference will be smaller,
but still appreciable for $M_{\rm pc}\ga 10^{-12} M_\odot$.
Such an explanation of the CCO radiation is compatible
only with the standard cooling scenario --- the difference of chemical
compositions could not account for a large ratio, $\sim 10$, of the
cap and surface temperatures required by the accelerated cooling.

Let us consider the hypothesis that the observed X-ray radiation 
is due to accretion onto a NS or a black hole (BH).
To provide a luminosity $L=L_{33}\times 10^{33}$~erg~s$^{-1}$,
the accretion rate should                                   
be $\dot{M}=L/(\xi c^2)= 1.1\times 10^{12} L_{33} \xi^{-1}$~g~s$^{-1}$,
where $\xi$ is the accretion efficiency ($\xi=
0.2 M_{1.4}R_{6}^{-1}$ for accretion onto the surface of a NS).
Although the luminosity and the accretion rate are very small compared
to typical values observed in accreting binaries, they are too high to
be explained by accretion from 
circumstellar matter (CSM) ---
very high CSM densities
and/or low object velocities relative to the accreting medium are required. 
For instance, the Bondi formula,
$\dot{M}=4\pi G^2M^2\rho v^{-3}$, 
gives the following relation between the CSM baryon density $n$
and velocity $v=100 v_{100}$ km~s$^{-1}$: 
$n=8\times 10^3 v_{100}^3 (0.2/\xi) M_{1.4}^{-2} L_{33}$~cm$^{-3}$.
Even at $v_{100}=1$,
which is lower than a typical pulsar velocity,
the required density exceeds that expected in the Cas~A interiors by about
3--4 orders of magnitude, unless the NS or BH moves within a much denser
(and sufficiently cold) CSM concentration.
This estimate for $n$ can be considered as a lower limit because 
accretion onto a BH, or onto a NS in the propeller regime, is much
less efficient.

We cannot, however, exclude that CCO is accreting from a secondary
component in a close binary or from a fossil disk which remained
after the SN explosion.
We can rule out a massive secondary component ---
from the above-mentioned $R$ and $I$ limits, we estimate
$M_R\ga +8$, $M_I\ga +8$. We can also exclude
a persistent low-mass X-ray binary (LMXB) or a transient LMXB in outburst 
--- the object would have a much
higher X-ray luminosity than observed, and the accretion disk would
be much brighter in the IR-optical range
(van Paradijs \& McClintock 1995).
However, CCO might be
a compact object with a fossil disk,
or an LMXB with a dwarf secondary component, 
in a long-lasting quiescent state 
(e.g., an M5 dwarf with $M_{\rm bol}=+9.8$ would have
$I\simeq 24.7$, for the adopted distance and extinction).
An indirect indication that CCO could be a compact accreting object
is that its luminosity and spectrum resemble
those of LMXBs in quiescence, although we have not seen
variability inherent to such objects.
If the accreting object were a young NS, it would be hard to 
explain how the matter accretes onto the NS surface ---
a very low magnetic field and/or long rotation period, 
$P\ga 10^2 B_{12}^{6/7} L_{33}^{-3/7} R_6^{15/7} M_{1.4}^{-2/7}$~s,
would be required
for the accreting matter to penetrate the centrifugal barrier.
The criterion suggested by Rutledge et al.~(1999) to distinguish
between the NS and BH LMXBs in quiescence, based on fitting the
quiescent spectra with the light-element NS atmosphere models, 
favors the BH interpretation, although the applicability
of this criterion to a system much younger than classical LMXBs may be
questioned. On the other hand, 
in at least some of BH binaries optical radiation emitted
by the accretion flow was detected in quiescence
(e.g., Narayan, Barret \& McClintock 1997)
at a level exceeding the upper limit on the
CCO optical flux. 
Finally, one could speculate that the CCO progenitor was a binary
with an old NS, 
and this old NS has sufficiently slow
rotation and low magnetic field to permit accretion onto the NS surface
from a disk of matter captured in the aftermath of the SN explosion.
In this case, CCO could have properties of an accreting X-ray pulsar
with a low accretion rate.
(A similar model was proposed by Popov 1998 for the central source
of RCW 103, although he assumed accretion
from the ISM.)

To conclude, we cannot firmly establish the nature of CCO based on
the data available --- it can be either an isolated NS with hot spots
or a compact object (more likely, a BH) accreting from a fossil disk
or from a dwarf binary companion. Although the CCO spectrum and luminosity
strongly resemble those of other radio-quiet compact sources in SNRs,
these sources may not necessarily
represent a homogeneous group --- e.g., the central source of Kes 73
shows 11.7 s pulsations and remarkable stability, and was proposed
to be a magnetar (Gotthelf et
al.~1999), whereas the central source of RCW 103 shows long-term
variability and no pulsations (Gotthelf, Petre \& Vasisht 1999).
We favor the isolated NS interpretation of CCO
because it has not displayed any variability. Critical observations
to elucidate its nature include searching for periodic and aperiodic
variabilities,
deep IR imaging, and longer \chan\ ACIS observations  which would
provide more source quanta for the spectral analysis.

We are grateful to Norbert Schulz for providing the ACIS response
matrices, to Gordon Garmire, Leisa Townsley and George Chartas 
for their advices on the ACIS data reduction, and to 
Niel Brandt, Sergei Popov, and Jeremy Heyl
for useful discussions.
The \ros\ and \einst\ data were obtained through
the High Energy Astrophysics Science Archive Research Center Online Service,
provided by the NASA's Goddard Space Flight Center.
The work was partially supported through
NASA grants NAG5-6907 and NAG5-7017.
{}
\begin{figure}
\plotone{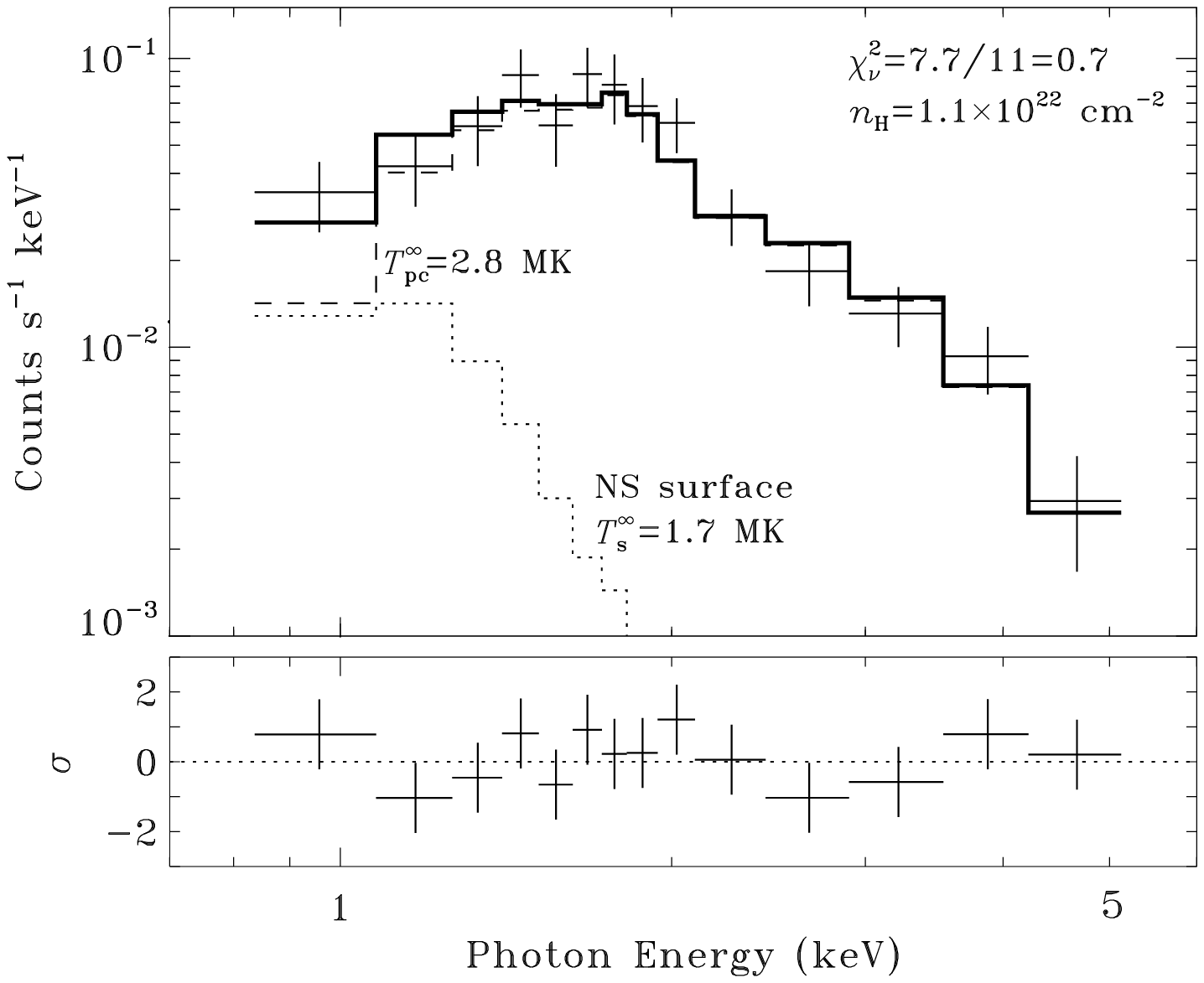}
\caption{
\chan\ ACIS-S3 countrate spectrum from the central compact
object of Cas A. The fit for hydrogen polar caps ($T_{\rm pc}^\infty=
2.8$ MK, $R_{\rm pc}=1$ km) on a cooler iron NS surface ($T_{\rm s}^\infty=
1.7$ MK, $R=10$ km) is shown.
}
\end{figure}
\begin{figure}
\vskip -2truein
\plotfiddle{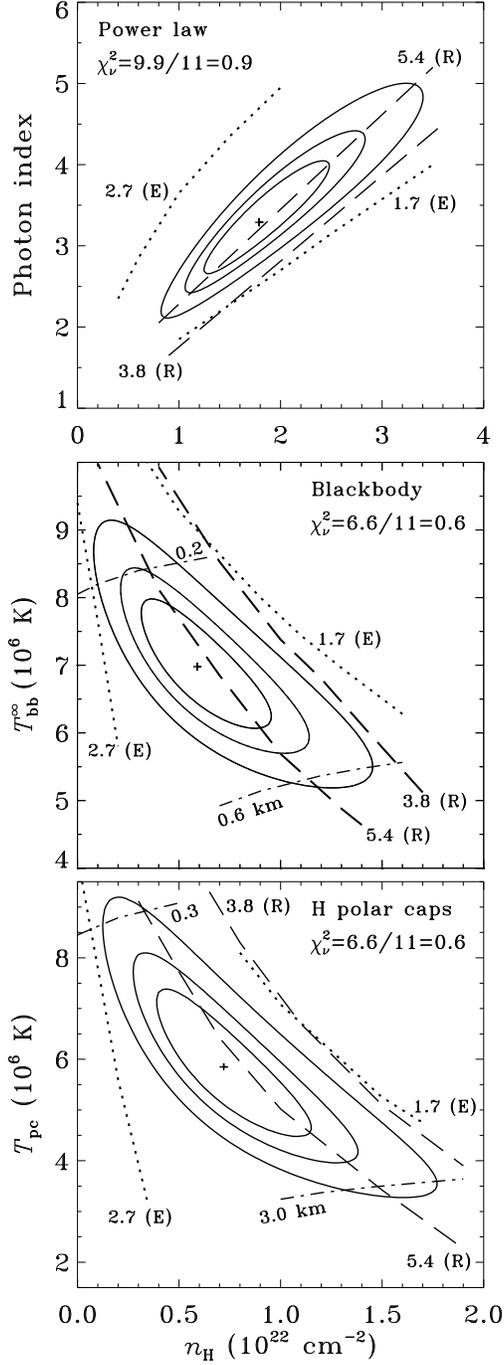}{8truein}{0}{90}{90}{-270}{-200}
\vspace*{1truein}
\caption{
67\%, 90\% and 99\% confidence regions obtained
from the fits to the spectrum of Figure 1, together with lines
of constant 
\ros\ HRI (long dashes)
and \einst\ HRI (dots) countrates, in counts per kilosecond. 
The dash-dot curves in the two lower panels are the
lines of constant radii of emitting areas (in km).
}
\end{figure}
\end{document}